\documentclass[usegraphicx,usenatbib]{mn2e}
\begin{document}

\title[XMM-Newton Discovery of an X-ray Transient]{XMM-Newton Discovery of the X-ray Transient XMMU~J181227.8$-$181234 in the Galactic Plane}

\author[Cackett et~al.]
 {Edward M. Cackett$^1$\thanks{emc14@st-andrews.ac.uk},
 Rudy Wijnands$^2$, Ron Remillard$^3$
\\ $^1$ School of Physics and Astronomy,
	University of St.~Andrews,
	KY16 9SS, Scotland, UK
\\ $^2$ Astronomical Institute `Anton Pannekoek',
        University of Amsterdam, Kruislaan 403, 1098 SJ,
        Amsterdam, the Netherlands
\\ $^3$ MIT Kavli Institute for Astrophysics and Space Research,
        77 Massachusetts Avenue, Cambridge, MA 02139, USA
}
\date{Received ; in original form }
\maketitle

\begin{abstract}
We report the discovery of an X-ray transient, observed in outburst with {\it XMM-Newton} on March 20, 2003, and with position $\alpha$ = 18h12m27.8s, $\delta$ = $-$18$^{\circ}$12$^{\prime}$34$^{\prime\prime}$ (J2000, approximate positional error 2$^{\prime\prime}$).  No known source is present at this position and the source was not detected during published {\it ROSAT} or {\it ASCA} observations of that region.  However,  the source may be associated with 1H1812-182 detected by {\it HEAO 1}, although the error bars on the {\it HEAO 1} position are very large and the two sources could also be unrelated. Therefore, we name the source XMMU~J181227.8$-$181234. Initially, the source was not detected using the All-Sky Monitor (ASM) on-board the {\it Rossi X-ray Timing Explorer}, however, reprocessing of the ASM data shows that the source was in fact detected and it was active for about 50 days. The X-ray spectrum of this transient is fitted equally well by an absorbed power-law (with a spectral index of 2.5) or multi-colour disk blackbody model (with kT$\sim$2 keV), where we find that the source is highly absorbed. We detect an unabsorbed 0.5-10 keV flux in the range $(2-5)\times10^{-9}$ ergs cm$^{-2}$ s$^{-1}$, which at a distance of 8 kpc corresponds to a 0.5-10 keV luminosity of $(1-4)\times10^{37}$ ergs s$^{-1}$. No pulsations were detected by timing analysis.  A colour-colour diagram from ASM data of different accreting objects suggests that the transient is a high-mass X-ray binary, as is also suggested by the high absorption compared to the average interstellar value in the direction of the source.  However, the power-law spectral index is far more typical of a low-mass X-ray binary.  Thus, we are unable to conclusively identify the nature of the transient.  We also report on three sources first detected by the {\it ASCA} Galactic Plane Survey that are close to this transient.

\end{abstract}


\section{Introduction}
X-ray binaries are accreting neutron stars and black holes and can be divided into persistent and transient sources. The transient sources exhibit many orders of magnitude variation in their X-ray luminosity due to similar large fluctuations in the mass accretion rate onto the compact objects. Due to instrument limitations, most of our understanding of these systems is based on the properties and behaviour of the brightest systems ($L_x > 10^{36} - 10^{37}$ erg~s$^{-1}$). Although quite a few lower luminosity systems (with $L_x$ between $10^{34}$ and $10^{36}$ erg~s$^{-1}$) are known, they remain enigmatic as they have not be studied in the same systematic way as their brighter counterparts. We have initiated several programs to systematically investigate the lower luminosity accreting neutron stars and black holes. One of these programs is to search and study very faint X-ray transients (with peak luminosities of $10^{34}$ and $10^{36}$ erg~s$^{-1}$) near the Galactic centre \citep[see][]{wijnands06}. 

Another program we are performing is to identify the many X-ray sources which were found during several different previous X-ray surveys of the Galactic centre and the Galactic plane \citep[e.g.,][]{sugizaki01,sidoli01,sakano02}. In particular, the {\it ASCA} Galactic Plane Survey \citep{sugizaki01} detected 163 sources in the Galactic Plane, and those that are at a distance of 8 kpc approximately fall into the $L_x = 10^{34}$ to $10^{36}$ ergs~s$^{-1}$ range.  The majority of the sources remain unidentified due to the poor spatial resolution of the {\it ASCA} observations. We have started a detailed investigation of the archival {\it XMM-Newton} and {\it Chandra} data for observations of these {\it ASCA} sources. Such observations will lead to better positional accuracy which will be used to follow up the X-ray sources at optical and IR wavelengths to try to constrain the nature of the sources. We will also study the X-ray spectral and timing properties which will provide additional insight into the nature of these sources. Whilst studying the archival {\it XMM-Newton} observations, we discovered a rather bright X-ray transient. In this paper we report on this discovery, as well as discussing 3 sources first detected by the {\it ASCA} Galactic Plane Survey. In future work we will discuss in detail the results of our archival searches for the {\it ASCA} and other low-luminosity sources.

\section{XMM-Newton Observation} \label{sec:obs}
\begin{figure}
  \centering
  \includegraphics[width=8cm]{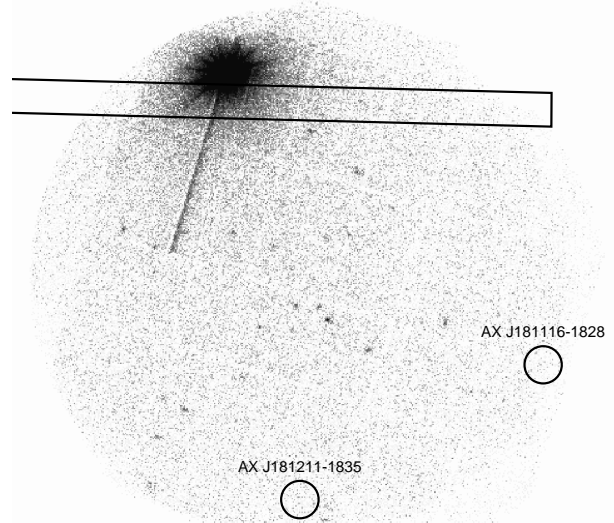}
  \caption{Combined EPIC image of the observed region.  The transient is clearly seen in the top left of the image.  The rectangular error box for source 1H1812-182 observed with the {\it HEAO 1} satellite is shown \citep{wood84}.  The error circles for sources  AX~J1811.2$-$1828 and AX J1812.1-1835 are shown which were detected during the {\it ASCA} Galactic Plane Survey \citep{sugizaki01}.}
  \label{fig:image}
\end{figure}

{\it XMM-Newton} \citep{jansen01} observed a region of the Galactic Plane in the direction of $\alpha$ = 18h12m6.3s, $\delta$ = $-$18$^{\circ}$22$^{\prime}$30$^{\prime\prime}$ on March 8, 2003 (see Fig.~\ref{fig:image}, ObsID 0152833701).  The observation exposure times are $\sim$8.7 ks for the EPIC-MOS and $\sim$7 ks for the EPIC-pn, no background flares are present.  The EPIC instrument was operated in full frame mode with the medium filter.  The Observation Data Files were processed to produce calibrated event lists using the {\it XMM-Newton} Science Analysis Software (SAS, version 6.5.0). 

In the field of view, we detect a bright source with position $\alpha$ = 18h12m27.8s, $\delta$ = $-$18$^{\circ}$12$^{\prime}$34$^{\prime\prime}$ (J2000). The statistical error in this source position is small (0\farcs05) as the source is bright, and so the positional error is dominated by the absolute astrometric accuracy of the telescope, which is estimated to be $\sim$$2^{\prime\prime}$ \citep{kirsch06}.  An X-ray source at this position is not identified in catalogues of previous \textit{ROSAT} and \textit{ASCA} surveys of the same region, identifying this source as being transient. However, this source lies close ($\sim$40\arcsec) to the 95\% confidence error box for the source 1H1812-182 detected during the NRL Large Area Sky Survey Experiment with the {\it HEAO 1} satellite \citep{wood84}.  Although the given position of 1H1812-182 ($\alpha$ = 18h15m27s, $\delta$ = $-$18$^{\circ}$12$^{\prime}$55$^{\prime\prime}$, J2000) is $\sim$0.7$^{\circ}$ from the position of the {\it XMM-Newton} source, it has a large extended error box with an approximate length of 2.1$^{\circ}$ and width of 0.03$^{\circ}$ (= 1.8\arcmin).  The {\it HEAO 1} source is therefore a possible X-ray detection of this {\it XMM-Newton} transient during a previous outburst, though we note that the large error box of 1H1812-182 does not rule out the possibility that it could well be another X-ray source that is unrelated.  1H1812-182 was detected with a flux of $0.1470 \pm 0.0025$ counts cm$^{-2}$ s$^{-1}$ in the 0.5-25 keV band which corresponds to a 2-10 keV flux of $7.0\pm0.1 \times 10^{-10}$ ergs cm$^{-2}$ s$^{-1}$ assuming a Crab-like spectrum \citep{wood84}.  

As this new {\it XMM-Newton} source has not been previously detected, we name it XMMU J181227.8$-$181234.  Initially, the source was not detected using the All-Sky Monitor (ASM) on-board the {\it Rossi X-ray Timing Explorer (RXTE)}, however, reprocessing of the ASM data showed that the source was in fact detected (see Section \ref{sec:rxte} for further details).

There is no optical counterpart identified in a Digitized Sky Survey (DSS) image and no counterpart in the 2MASS All-Sky Catalog of Point Sources \citep{cutri2003} or USNO-B1.0 \citep{monet2003} catalogue within 6\arcsec of the source position.  The large absorption in this direction makes finding an optical counterpart particularly hard.

\section{Timing and Spectral Analysis}

\begin{figure}
  \centering
  \includegraphics[angle=270, width=7.5cm]{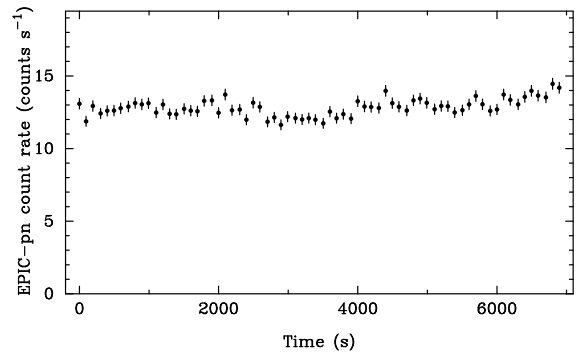}
  \caption{EPIC-pn 0.2-10 keV background subtracted lightcurve for XMMU J181227.8$-$181234.  Time binning is 100s.}
  \label{fig:pn_lc}
\end{figure}

Extracting a 0.2-10 keV lightcurve from the EPIC-pn and MOS detectors we detect no X-ray bursts (see Fig.~\ref{fig:pn_lc}).  We use a periodogram analysis \citep{lomb76,scargle82} of the combined EPIC-pn and MOS 0.2-10 keV lightcurve, with a time bin of 10s, to search for any pulsations which would help in identifying the nature of the source, but none were found.

This source has a count rate high enough to cause pile-up.  Therefore, to perform spectral analysis of this data the pile-up needs to be corrected for.  We extract the spectrum using two different methods.  Firstly, from an annulus around the source thus excluding the source core which is most significantly effected by pile-up, and secondly from the read-out streak present in the EPIC-pn data.  This second method provides a check as to whether the annulus spectrum is significantly affected by the energy-dependence of the point-spread function.

In the annulus method, we use an annulus of inner radius 15\arcsec~and outer radius 45\arcsec.  The background was extracted from a 3\arcmin~circular region close to the transient and containing no detected sources.  For the MOS cameras, only single-pixel events (pattern = 0) were selected as recommended for dealing with pile-up by \citet{molendi_sembay_03}.  We use the SAS tool EPATPLOT to make sure that no pile-up was present.  Response matrices and ancillary files were created using RMFGEN and ARFGEN.

\begin{figure}
  \centering
  \includegraphics[width=7cm]{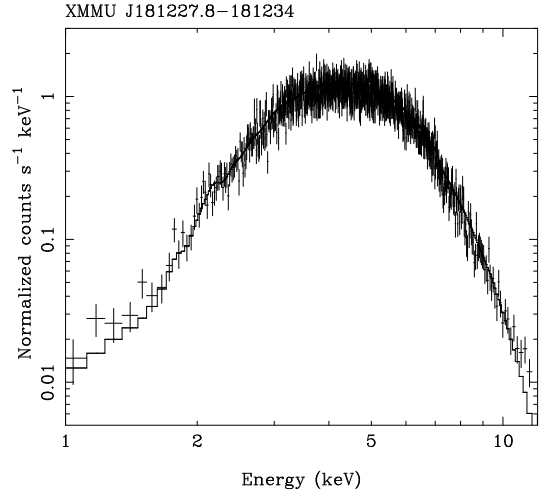}
  \caption{EPIC-pn spectrum for XMMU J181227.8$-$181234 with best-fitting absorbed multi-colour disk blackbody model.  The spectrum was extracted from an annulus around source to reduce the effects of pile-up.}
  \label{fig:pn_spec}
\end{figure}

The data were binned to have a minimum of 25 counts per bin. We use XSPEC (v.11) \citep{arnaud96} to fit the EPIC-pn, MOS1 and MOS2 data simultaneously. Two different models were fit to the data - an absorbed power-law model and an absorbed multi-colour disk blackbody model (diskbb). Table~\ref{tab:fits} gives the best-fitting parameters for both models and the best-fitting multi-colour disk blackbody model is shown if Fig.~\ref{fig:pn_spec}.  We find that both models give reasonably good fits.  From the spectrum and model fits it is apparent that the source is highly absorbed suggesting that it is not a nearby lower-luminosity source.  The power-law fit gives a spectral index, $\Gamma = 2.47 \pm 0.05$, where as the multi-colour disk blackbody fit gives quite a high characteristic temperature of $kT = 2.14 \pm 0.04$ keV.  These model fits lead to an unabsorbed 0.5-10 keV flux of $5.3 \pm 0.4 \times 10^{-9}$ ergs cm$^{-2}$ s$^{-1}$ and $1.81 \pm 0.02\times 10^{-9}$ ergs cm$^{-2}$ s$^{-1}$ for the power-law and multi-colour disk blackbody models respectively.  Assuming that this source is at a distance of 8 kpc, this corresponds to a 0.5-10 keV luminosity of $4.1 \pm 0.3 \times 10^{37}$ ergs cm$^{-2}$ s$^{-1}$ for the power-law model and $1.4 \pm 0.02 \times 10^{37}$ ergs cm$^{-2}$ s$^{-1}$ for the multi-colour disk blackbody model, quite typical of other X-ray transients. 

The readout streak is only significantly present in the pn image, thus we only extract the readout streak spectrum from the pn data.  Binning the data to have a minimum of 25 counts per bin, we again fit in XSPEC using both an absorbed multi-colour disk blackbody model and an absorbed power-law model.  The best-fitting parameters are given in Table~\ref{tab:fits}, and are seen to be consistent with the parameters from the annulus spectrum.  We do not give fluxes for the readout streak spectrum as the photons only hit the CCD in the streak region when the CCD is being read-out, and so the exposure time is incorrect.

\begin{table*}
\begin{center}
\caption{Model fits to the X-ray spectrum of XMMU J181227.8$-$181234 when fitting the EPIC-pn, MOS1 and MOS2 data simultaneously to a spectrum extracted from the annulus, and when fitting the EPIC-pn spectrum extracted from the readout streak.  1-$\sigma$ errors on the parameters are given.}
\label{tab:fits}
  \begin{tabular}{lcccc}
    \hline
     & \multicolumn{2}{c}{Annulus} & \multicolumn{2}{c}{Readout Streak}  \\
    Parameter & Power-law & diskbb & Power-law & diskbb \\
    \hline
    $N_H$ ($\times10^{22}$ cm$^{-2}$) & $12.8 \pm 0.3$ & $9.7 \pm 0.2$ 
    & $14.3^{+1.6}_{-1.4}$ & $10.8^{+1.0}_{-0.9}$ \\
    Spectral Index & $2.47 \pm 0.05$ & & $2.34^{+0.23}_{-0.22}$ &\\
    kT (keV) &  &  $2.14 \pm 0.04$ & & $2.38^{+0.25}_{-0.22}$\\
    0.5-10 keV absorbed flux  & 
    $7.7 \pm 0.1$ & $7.6 \pm 0.1 $ & &\\
    ($\times 10^{-10}$ ergs cm$^{-2}$ s$^{-1}$) & & & &\\
    0.5-10 keV unabsorbed flux  & 
    $5.3 \pm 0.4$ & $1.81 \pm 0.02 $\\
    ($\times 10^{-9}$ ergs cm$^{-2}$ s$^{-1}$) & & & &\\
    $\chi_{\nu}^{2}$ (d.o.f) & 1.4 (1459) & 1.3 (1458) & 0.99 (124)
     & 0.90 (124)\\
    \hline
  \end{tabular}
\end{center}
\end{table*}

\section{{\it RXTE} All Sky Monitor observations}\label{sec:rxte}
\begin{figure}
  \centering
  \includegraphics[width=8cm]{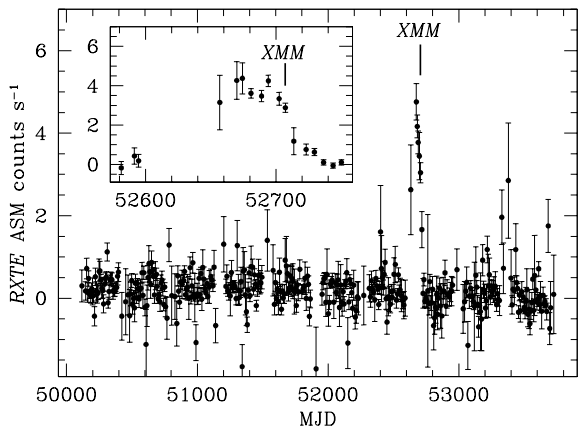}
  \caption{7-day averaged {\it RXTE} ASM lightcurve for XMMU~J181227.8$-$181234, with the time of the {\it XMM-Newton} observation marked.  The inset shows the outburst in more detail, again with the time of the {\it XMM-Newton} observation marked.  There is clearly only one outburst detected during the lifetime of {\it RXTE}.}
  \label{fig:asm}
\end{figure}

Initially, the detection of this transient with the {\it RXTE} ASM was missed, due to the mask that is put around bright sources when looking at deep sky maps.  In this case the bright persistent low mass X-ray binary GX~13+1 (4U~1811-17) is $\sim$1.2${^{\circ}}$  away from the transient.  Given the position and time of the outburst, we detect the transient with the {\it RXTE} ASM. We have also derived the source position from first principles, given the outburst time interval, that is consistent with the {\it XMM-Newton} position.  Figure \ref{fig:asm} shows the {\it RXTE} ASM lightcurve for this source with the outburst clearly visible.  The outburst lasts for approximately 50 days, at least, and there is only one outburst detected during the lifetime of {\it RXTE}.  The {\it XMM-Newton} observation occurs after the peak of the outburst.  The ASM count rate at the time of the {\it XMM-Newton} observation is approximately 3 counts s$^{-1}$. Using PIMMS, and assuming this count rate and the best-fitting power-law spectral model from Table~\ref{tab:fits}, this corresponds to an unabsorbed 0.5-10 keV flux of 6~$\times 10^{-9}$ ergs cm$^{-2}$ s$^{-1}$, similar to that detected with {\it XMM-Newton}.

The coarse shape of the spectrum in X-ray binary systems may help to reveal the type of the source. To investigate this question, we use the hardness ratios derived from the source count rates in the three standard ASM energy bands: $A$ (1.5-3 keV), $B$ (3-5 keV) and $C$ (5-12 keV). The hardness ratios are then defined as HR1 = $B/A$ and HR2 = $C/B$. We expect that HR1 is sensitive to both the spectral shape of the X-ray source and the amount of interstellar absorption, while HR2 is more sensitive to the spectral shape.  

The `dwell-by-dwell' data for a particular source in the ASM archive provides independent measurements from each 90 s exposure by one of the three ASM cameras.  We select cameras 2 and 3 only for this exercise.  Camera 1 is excluded because it has a secular gain drift, caused by a pinhole Xe leak discovered before {\it RXTE} launch, and it is rigidly calibrated to the other cameras for the specific spectral shape of the Crab Nebula.  For other spectral shapes, the hardness ratios from camera 1 exhibit a time dependence that would add systematic noise in the present study.

For persistent sources (6 atolls, 6 Z sources, 4 classical X-ray pulsars, and the 2 LMC black holes), we compute the ASM count rates (weighted mean) in each energy band over a long time interval, arbitrarily chosen to be MJD 51000 -- 52000.  We also sample the persistent black hole binary Cyg X--1 in each of two X-ray states: soft (MJD 52250--52450) and hard (MJD 50720 -- 50920).  Count rates are determined for X-ray transients (8 atolls, 6 classical X-ray pulsars, and 7 black holes) for the 7-day time interval that captures the peak of the outburst. This is evaluated by finding the maximum flux in the weighted mean, while sliding a 7-day window along the light curve sampled in 1-day bins.

\begin{figure}
  \centering
  \includegraphics[width=7cm]{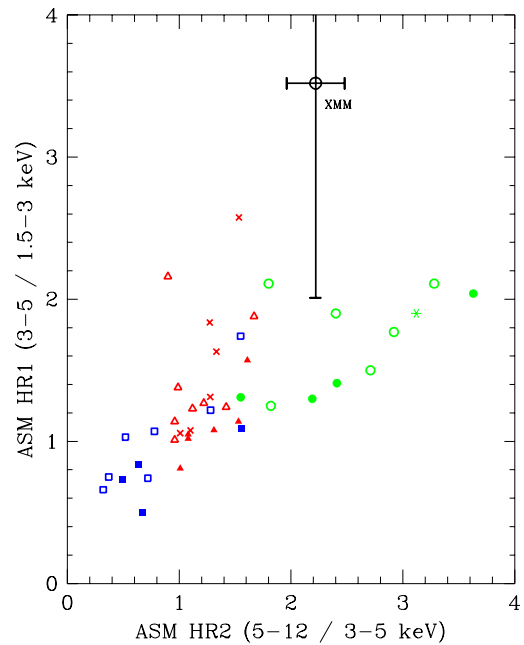}
  \caption{{\it RXTE} ASM colour-colour diagram. The shape (and also the colour) denote source types, an open symbol is a transient; filled symbols are persistent sources.  The squares (blue) are black-hole binaries with two states for Cyg X-1.  Triangles (red) denote the bursters.  Crosses (red) denote the Z-sources (all of these are persistent).  Circles (green) denote the pulsars.  The green asterisk is the high-mass X-ray binary 4U 1700-377.}
  \label{fig:asmtypes}
\end{figure}

From this analysis, we find that a colour-colour diagram using these two ASM hardness ratios can give a good indication of source type (see Fig.~\ref{fig:asmtypes}).  We note that HR1 is quite uncertain for this transient due to the low number of counts in the 1.5-3 keV band caused by the high absorption towards this object.  The HR2 ratio, however, is less susceptible to interstellar absorption.  We find that HR2 = 2.45, suggesting that this source is a high-mass X-ray binary pulsar.

\section{Discussion}
\subsection{The nature of XMMU~J181227.8$-$181234}
From our spectral fits to XMMU~J181227.8$-$181234 we find a high absorption towards this source.  Using the FTOOL nH \citep{dickeylockman90}, we get $N_H \sim 2\times 10^{22}$ cm$^{-2}$ in the direction of this source - a factor at least 5 different from the value from spectral fitting.  This may suggest that the source has significant intrinsic absorption which would be indicative of a high-mass X-ray binary (HMXB). INTEGRAL observations of the Galactic plane have discovered sources with high intrinsic absorption that are likely HMXBs \citep[e.g.,][]{patel04,walter04,lutovinov05a,lutovinov05b,bodaghee06}.  The intrinsic absorption is due to the accretion process in the majority of HMXBs in which the compact objects accrete matter from the stellar wind of a young companion star. The stellar wind can cause significant absorption.  In low-mass X-ray binaries (LMXBs), however, accretion occurs via Roche-lobe overflow through the inner Lagrangian point, thus in the majority of cases no material is available to cause significant absorption.  Only in high inclination systems significant internal absorption might be present. However, for this transient we do not find any of the characteristic count rate variations (i.e., dips and eclipses) which would indicate a high inclination.

HMXB pulsars typically show significant slow pulsations, however, there are notable exceptions, for instance the well known HMXB, 4U 1700-377 (marked in Fig. \ref{fig:asmtypes} as an asterisk), has no detectable pulsations.
In addition, HMXBs can typically be fitted by an absorbed power-law with spectral index, $\Gamma \sim 1$.  There are some lower-luminosity HMXBs that have spectra that are softer \citep[with $\Gamma \sim 2-2.5$, see, e.g.][]{reigroche99}, though these sources are persistent and have $L_x \sim 10^{34}$ ergs s$^{-1}$, where as XMMU~J181227.8$-$181234 is transient in nature and is significantly brighter (with $L_x$ a few times 10$^{37}$ ergs s$^{-1}$). XMMU~J181227.8$-$181234 has no pulsations detected in the lightcurve and it is fit by a power-law spectral index of $\Gamma = 2.5$ which is far more typical of LMXBs. Thus, although the ASM colours and possible intrinsic absorption suggest that this source may be a HMXB, the power-law index is more suggestive of a LMXB.  We are therefore unable to conclusively determine whether this transient is either a HMXB or a LMXB, and an IR follow-up is needed to search for a counterpart at this wavelength.  As no optical or IR counterpart was found with USNO-B1.0 or 2MASS this suggests that the source is not a nearby lower-luminosity object which is supported by the high measured $N_H$.

\subsection{{\it ASCA} Galactic Plane Survey sources}

The {\it ASCA} Galactic Plane Survey \citep{sugizaki01} detected 163 discrete sources in the 0.7-10 keV energy band within the central region of the Galactic plane with a spatial resolution of $\sim$3\arcmin~and a positional error of $\sim$1\arcmin. The limiting flux of the survey in the 0.7-10 keV band was $\sim$$3\times10^{-13}$ ergs cm$^{-2}$ s$^{-1}$ which corresponds to a luminosity of $\sim$$2\times10^{33}$ ergs s$^{-1}$ assuming a distance of 8~kpc.  We use PIMMS to predict the EPIC-pn on-axis count rate assuming such a flux, a column density $N_H = 2 \times 10^{22}$ cm$^{-2}$, and with both a power-law with spectral index, $\Gamma = 0$ and spectral index, $\Gamma = 3$.  We calculate the EPIC-pn on-axis count rate for this to be approximately 0.014 counts s$^{-1}$ when $\Gamma = 0$ and 0.041 counts s$^{-1}$ when $\Gamma = 3$.  Thus, for an exposure time of 7 ks, the lower limit of {\it ASCA} Galactic Plane Survey corresponds roughly to between 100 and 290 counts on-axis (0.5-10 keV) for typical spectra.  We note that off-axis the effective exposure time drops rapidly due to vignetting.  By around 15$^{\prime}$ off-axis, the vignetting factor is about 0.3 (see Fig. 13 of the {\it XMM-Newton} User's Handbook\footnote{available from http://xmm.vilspa.esa.es}), thus the lower limit drops to between 30 and 90 counts.  Thus we expect to detect any persistent point sources from the {\it ASCA} Galactic Plane Survey with this {\it XMM-Newton} observation. 

Two of the sources (AX~J1811.2$-$1828 and AX~J1812.1$-$1835) from this survey lie within the field of view of this {\it XMM-Newton} observation and another of the sources (AX~J1812.2$-$1842) is in an {\it XMM-Newton} observation of a nearby region. We discuss the {\it XMM-Newton} observation of these sources here.

\subsection{AX~J1811.2$-$1828}
AX~J1811.2$-$1828 was detected for the first time with {\it ASCA} \citep{sugizaki01}.  No spectral model was fit to data from AX~J1811.2$-$1828, though these authors quote a 2-10 keV count rate of 6.3 counts ks$^{-1}$ GIS$^{-1}$.  Depending on the spectral model, this converts to a 2-10 keV flux between $(4 - 11.5) \times 10^{-13}$ ergs cm$^{-2}$ s$^{-1}$.

\begin{figure}
  \centering
  \includegraphics[width=8.5cm]{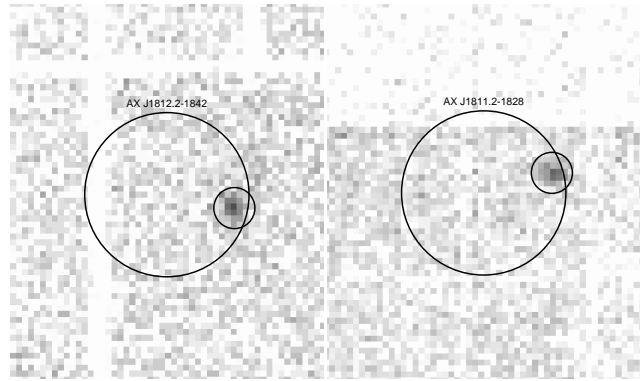}
  \caption{The two {\it ASCA} sources AX~J1812.2$-$1842 (left panel) and AX~J1811.2$-$1828 (right panel) are clearly detected with {\it XMM-Newton} with a much improved source position.  The larger circles are the {\it ASCA} error circles for these sources, whilst the smaller circles are the 15$^{\prime\prime}$ source extraction regions we use here.}
  \label{fig:asca_sources}
\end{figure}

The majority of the error circle for this source falls off the edge of the MOS1 and MOS2 detectors, and so we can only use the pn detector to search for a detection in this {\it XMM-Newton} observation.  No bright source is detected within the error circle, however, we tentatively detect a possible faint source on the edge of the error circle for this source with $40 \pm 9$ net counts in the 0.2-12 keV range for the pn detector.  To check whether this source is in fact real, we analyse two additional {\it XMM-Newton} archival observations (ObsID 0152834201 and 0152835401) in the region of this source.  In both observations we detect a source with a position $\alpha$ = 18h11m13.3s, $\delta$ = $-$18$^{\circ}$27$^{\prime}$57$^{\prime\prime}$ (J2000), and a positional error of 2\farcs2 (including both statistical and boresight errors).  The position of the potential source in the original data is consistent with that of the source in the two extra data sets and thus the source is detected in all three {\it XMM-Newton} observations.  In Fig. \ref{fig:asca_sources} the improved position of this source can be seen in a combined image from all 3 {\it XMM-Newton} observations. In ObsID 0152835401 the source is on the very edge of the field-of-view and so we only analyse the spectrum from the ObsID 0152834201 data, where the source is bright enough to bin the spectrum with enough counts to validly use $\chi^2$ statistics in the spectral fitting.

We extract the spectrum from the MOS1, MOS2 and pn detectors using a circular source region with radius 15$^{\prime\prime}$, and we extract the background from an annulus around the source with inner and outer radii of 1$^{\prime}$ and 2$^{\prime}$ respectively. We bin the spectrum to 15 counts per bin, and fit the spectra simultaneously with an absorbed power-law model using XSPEC.  The best-fitting model has parameters $N_H = 0.6^{+0.6}_{-0.3} \times 10^{22}$ cm$^{-2}$ and $\Gamma = 1.5^{+0.7}_{-0.3}$.  The reduced-$\chi^2$ for this fit is 0.8.  For comparison with the {\it ASCA} detection we calculate the 2-10 keV flux, which we find to be $2.8^{+0.6}_{-0.4} \times 10^{-13}$ ergs cm$^{-2}$ s$^{-1}$ and the unabsorbed 2-10 keV flux is $2.9 \pm 0.6 \times 10^{-13}$ ergs cm$^{-2}$ s$^{-1}$.  This flux is slightly lower than the flux detected by {\it ASCA}.  From the spectral fits this source has a reddening smaller than the interstellar absorption in this direction, thus it is likely closer than a distance of 8~kpc.  As the distance to this source is unknown it is hard to determine what type of source this is with just this spectral fits, and the power-law index does not rule out a HMXB, LMXB or cataclysmic variable. 
We conclude that we have detected AX~J1811.2$-$1828 with {\it XMM-Newton} with a 2-10 keV flux which was slightly lower than that seen by {\it ASCA}.

We find several possible optical and IR counterparts to this X-ray source in the USNO-B1.0 and 2MASS catalogues.   The USNO-B1.0 source named
USNO-B1.0~0715-0555562 is 1\farcs8 away and has a B2 magnitude of 19.8 and a R2 magnitude of 18.0.  USNO-B1.0~0715-0555563 is 2\farcs6 away, and has B2 and R2 magnitudes of 19.3 and 16.5 respectively.  USNO-B1.0~0715-0555571 is 2\farcs7 away and has B2 and R2 magnitudes of 19.11 and 15.35 respectively.  The 2MASS source named 2MASS~J18111351-1827565  is a distance of 3\farcs1 away from the X-ray source and has J, H and K magnitudes of  12.3, 11.2, and  10.9 respectively.  Optical/IR follow-up spectroscopy of these sources is needed to determine the spectral type, and hence the likely nature of this source.

\subsection{AX~J1812.1$-$1835}
\citet{sugizaki01} identify AX~J1812.1$-$1835 with the supernova remnant G12.0-0.1 \citep[$\alpha$ = 18h11m53.3s, $\delta$ = $-$18$^{\circ}$35$^{\prime}$54$^{\prime\prime}$,][]{clark75,altenhoff79,kassim92,dubner93,green97,green04}.  Their spectral fits to AX~J1812.1$-$1835 lead to a flux for the source of $\sim$$9\times10^{-13}$ ergs cm$^{-2}$ s$^{-1}$ in the 0.7-10 keV band.  No source is detected within the error circle of AX~J1812.1$-$1835.  To determine an upper-limit for this source, we extract the count rate from the faintest source detected at a similar off-axis angle.  Using an extraction radius of 15$^{\prime\prime}$ for a source located at $\alpha$ = 18h12m43.7s, $\delta$ = $-$18$^{\circ}$31$^{\prime}$58$^{\prime\prime}$ (J2000) we get $51 \pm 10$ background-subtracted counts in the 0.2-12 keV range for the pn detector.  A source-free annulus between 40$^{\prime\prime}$ and 100$^{\prime\prime}$ was used to determine the background counts.  The average effective exposure in the source extraction region is 2429s, giving a net count rate of $0.021 \pm 0.004$ counts s$^{-1}$.  The average interstellar absorption in the direction of this source is $2 \times 10^{22}$ cm$^{-2}$.  Using PIMMS and assuming this value we get a 0.5-10 keV flux for the source in the range $(1.5 - 4.2) \times 10^{-13}$ ergs cm$^{-2}$ s$^{-1}$ depending on the power-law spectral index which we vary from 0 to 3.  Thus, if AX~J1812.1$-$1835 is a point source, and has not varied in flux since the {\it ASCA} observation it should be detected, which it is not.

There are several possible explanations for the lack of detection.  Firstly, if this source is not associated with the supernova remnant (SNR), then the flux from this source has dropped by a factor of at least 2 between the {\it ASCA} and {\it XMM-Newton} observations, identifying the source as variable.  However, it is likely that this source is associated with the SNR G12.0-0.1.  Studying the VLA 1465~MHz radio observation of this SNR
\citep[see Fig. 3 from][]{dubner93} we find that although the bright thermal source is located several arcminutes away from AX~J1812.1$-$1835, the {\it ASCA} source does lie within the incomplete shell to the east.  If AX~J1812.1$-$1835 is associated with this SNR, then the emission will be diffuse, and therefore below our detection limit.

\subsection{AX~J1812.2$-$1842}
AX~J1812.2$-$1842 was first detected by the {\it ASCA} Galactic Plane Survey.  No spectral model was fit to data from AX~J1812.2$-$1842 by \citet{sugizaki01}, though these authors quote a 2-10 keV count rate of 5.4 counts ks$^{-1}$ GIS$^{-1}$.  Depending on the spectral model, this converts to a 2-10 keV flux between $(3 - 10) \times 10^{-13}$ ergs cm$^{-2}$ s$^{-1}$.

Although not in the field-of-view of the {\it XMM-Newton} observation containing the transient, it is in the data with ObsID 0152835401.  The source is detected with a position $\alpha$ = 18h12m10.5s, $\delta$ = $-$18$^{\circ}$42$^{\prime}$10$^{\prime\prime}$ (J2000), and a positional error of 2\farcs3 (including both statistical and boresight errors).  The improved source position can be seen in a combined MOS1, MOS2 and pn image (Fig. \ref{fig:asca_sources}, left panel).

We extract the spectrum from the MOS1, MOS2 and pn detectors using a circular source region with radius 15$^{\prime\prime}$, and we extract the background from an annulus around the source with inner and outer radii of 1$^{\prime}$ and 2$^{\prime}$ respectively. We bin the spectrum to 15 counts per bin for the MOS1 and MOS2 data and 40 counts per bin for the pn data. The spectra are fit simultaneously with an absorbed power-law model using XSPEC.  The best-fitting model has parameters $N_H = 0.5^{+1.3}_{-0.5} \times 10^{22}$ cm$^{-2}$ and $\Gamma = 0.0^{+0.7}_{-0.5}$.  The reduced-$\chi^2$ for this fit is 0.7.  For comparison with the {\it ASCA} detection we calculate the 2-10 keV flux, which we find to be $9.4^{+2.0}_{-1.7} \times 10^{-13}$ ergs cm$^{-2}$ s$^{-1}$ and the unabsorbed 2-10 keV flux is $9.6^{+1.8}_{-1.2} \times 10^{-13}$ ergs cm$^{-2}$ s$^{-1}$.  This flux is consistent with the source flux detected by {\it ASCA}.  The low $N_H$ suggests that the source maybe closer than 8~kpc, however, we note the large uncertainty in this parameter.  The power-law spectral index suggests that this source could be a cataclysmic variable, though we cannot be conclusive.

A search for optical and IR counterparts finds no sources within the errors for the position.  The nearest optical source, USNO-B1.0~0712-0536743, is a distance of 3\farcs5 away, and has B2 and R2 magnitudes of 18.4 and 17.8 respectively.  The nearest IR source, 2MASS~J18121061-1842133, is a distance of 4\farcs3, and has J, H and K magnitudes of 14.8, 12.3 and 10.9 respectively.

\section{Conclusions} \label{sec:conc}

We have discovered an X-ray transient with position $\alpha$ = 18h12m27.8s, $\delta$ = $-$18$^{\circ}$12$^{\prime}$34$^{\prime\prime}$ (J2000) detected by {\it XMM-Newton} and the {\it RXTE} ASM.  This source was not previously detected by either {\it ROSAT} or {\it ASCA}, though it may be associated with 1H1812-182 detected by {\it HEAO 1} which has a large extended error box that is close to the transient source position. The X-ray spectrum of this transient is fitted equally well by an absorbed power-law or multi-colour disk blackbody model, where we find that the source is highly absorbed.  We detect an unabsorbed 0.5-10 keV flux in the range $(2-5)\times10^{-9}$ ergs cm$^{-2}$ s$^{-1}$, which at a distance of 8 kpc corresponds to a 0.5-10 keV luminosity of $(1-4)\times10^{37}$ ergs s$^{-1}$, quite typical of other Galactic X-ray binary transients. Using a periodogram analysis we searched for pulsations to help with source identification, but none were found.  The high column density to the source and lack of optical/IR counterpart in the USNO-B1.0 and 2MASS catalogues suggests that the source is not a nearby source with lower luminosity.  
A colour-colour diagram from {\it RXTE} ASM data of different accreting objects suggests that the transient is a high-mass X-ray binary, as is also suggested by the high absorption compared to the average interstellar value in the direction of the source.  However, the power-law spectral index is far more typical of a low-mass X-ray binary.  Thus, we are unable to conclusively identify the nature of the transient.  Follow-up IR observations are required to help determine the nature of this source.

We also report on three sources first detected by the {\it ASCA} Galactic Plane Survey that lie within the field-of-view of this observation.  We do not detect AX~J1812.1$-$1835, though this is likely because it is associated with the supernova remnant G12.0-0.1 and so the diffuse emission would not be bright enough to be detected with this observation.  For AX~J1811.2$-$1828, we detect a source towards the edge of the error circle with a 2-10 keV flux of $2.8^{+0.6}_{-0.4} \times 10^{-13}$ ergs cm$^{-2}$ s$^{-1}$, slightly lower than the {\it ASCA} observation of this source.  The spectrum is fit with an absorbed power-law with $N_H = 0.6^{+0.6}_{-0.3} \times 10^{22}$ cm$^{-2}$ and $\Gamma = 1.5^{+0.7}_{-0.3}$.  We also detect AX~J1812.2$-$1842 with a 2-10 keV flux of $9.4^{+2.0}_{-1.7} \times 10^{-13}$ ergs cm$^{-2}$ s$^{-1}$, consistent with the {\it ASCA} observation of this source.  The spectrum is fit with an absorbed power-law with $N_H = 0.5^{+1.3}_{-0.5} \times 10^{22}$ cm$^{-2}$ and $\Gamma = 0.0^{+0.7}_{-0.5}$.  We cannot conclusively determine the nature of AX~J1811.2$-$1828 or AX~J1812.2$-$1842 with these observations and optical/IR follow-ups are needed.  Many sources from the {\it ASCA} Galactic Plane Survey remain unidentified in nature, and require further X-ray observations with {\it XMM-Newton} and {\it Chandra} to provide sub-arcsecond source positions allowing for optical/IR follow-up to aid with source identification as well as providing further X-ray temporal and spectral information.

\subsection*{Acknowledgements}

EMC gratefully acknowledges the support of PPARC.  {\it XMM-Newton} is an ESA science mission with instruments and contributions directly funded by ESA Member States and NASA.

\bibliographystyle{mn2e}
\bibliography{iau_journals,qNS_mnras}

\end{document}